\documentclass[pdftex,twocolumn,epjc3]{svjour3}       

\RequirePackage[T1]{fontenc}

\smartqed  

\RequirePackage{graphicx}
\RequirePackage{mathptmx}      
\RequirePackage{flushend}
\RequirePackage[numbers,sort&compress]{natbib}
\RequirePackage[colorlinks,citecolor=blue,urlcolor=blue,linkcolor=blue]{hyperref}

\journalname{Eur. Phys. J. C}

\begin{document}


\title{Top quark mass measurements at and above threshold at CLIC}

\author{Katja Seidel\thanksref{addr1}
        \and
        Frank Simon\thanksref{e1,addr1}
        \and
        Michal Tesa\v{r}\thanksref{addr1}
        \and
        Stephane Poss\thanksref{addr2} 
}

\thankstext{e1}{e-mail: fsimon@mpp.mpg.de}

\institute{Max-Planck-Institut f\"ur Physik, F\"ohringer Ring 6, 80805 Munich, Germany \label{addr1}
          \and
          CERN, 1211 Geneva, Switzerland \label{addr2}}

\date{Received: 15 March 2013 / Revised: 26 July 2013}

\maketitle

\begin{abstract}
We present a study of the expected precision of the top quark mass determination, measured at a linear $e^+e^-$ collider based on CLIC technology. GEANT4-based detector simulation and full event reconstruction including realistic physics and beam-in\-duced background levels are used. Two different techniques to measure the top mass are studied: The direct reconstruction of the invariant mass of the top quark decay products and the measurement of the mass together with the strong coupling constant in a threshold scan, in both cases including first studies of expected systematic uncertainties. For the direct reconstruction, experimental uncertainties around 100 MeV are achieved, which are at present not matched by a theoretical understanding on a similar level. With a threshold scan, total uncertainties of around 100 MeV are achieved, including theoretical uncertainties in a well-defined top mass scheme. For the threshold scan, the precision at ILC is also studied to provide a comparison of the two linear collider technologies.
\end{abstract}

\section{Introduction}
\label{sec:Intro}

As the heaviest Standard Model particle, the top quark is of particular interest since it couples most strongly  to the Higgs field and may provide sensitivity to Beyond the Standard Model physics. Its mass is a key parameter providing, together with the Higgs mass, constraints on the stability limit of the Standard Model \cite{Isidori:2001bm}. Currently, the uncertainty of the top quark mass is the leading uncertainty in this evaluation \cite{Degrassi:2012ry}. At the Tevatron, the top mass has now been determined with a combined precision of 1 GeV to $173.18  \pm 0.56\, \mathrm{(stat.)} \pm 0.75\,  \mathrm{(syst.)}$ GeV \cite{Aaltonen:2012ra}, and the combined LHC results at 7 TeV reach a total error of 1.4 GeV with $173.3 \pm 0.5\, \mathrm{(stat.)} \pm 1.3\,  \mathrm{(syst.)}$ GeV \cite{LHCTopCombination}. In both cases, the largest uncertainties are systematic, which are expected to improve only slowly as the analyses progress. 

Experimentally, the top quark offers the unique possibility to study an almost free quark, since its lifetime is considerably shorter than the QCD hadronization time. Due to its large mass, it is also the only quark so far not studied in electron-positron annihilation. Experiments at a future $e^+e^-$ collider at the high energy frontier offer the possibility for a wide variety of studies involving top quarks, ranging from the precise measurement of the top quark mass and width to the investigation of asymmetries, providing large sensitivity to various New Physics models \cite{Brau:2012hv}.

At such a collider, the mass of the top quark can be determined with two different  techniques: through the direct reconstruction of top quarks from their decay products at energies above the production threshold, and through a scan of the top-pair production threshold. The experimentally well-defined invariant mass is interpreted by comparing the measured distribution to that predicted by simulations. This provides the result in the context of the used event generator. The generator mass is not well-defined theoretically, and non-perturbative corrections could be substantial. Progress has been made recently in establishing connections between the top mass parameters used in theory and experimentally accessible parameters \cite{Fleming:2007qr, Fleming:2007xt}, which provide the potential for theoretical uncertainties below a scale of $\Lambda_{\mathrm{QCD}}$. The observables used in  \cite{Fleming:2007qr, Fleming:2007xt} do not match the invariant mass definition used for the top mass determination above threshold, which was chosen to allow the use kinematic constraints to improve the overall event reconstruction and to provide the possibility of beam-related background suppression via particle-level cuts and jet finding. This introduces additional, potentially sizeable, uncertainties in the theoretical interpretation of the measured value. Given the influence of non-perturbartive effects on the invariant mass, it is at present not clear if a sufficient improvement in the theoretical understanding is possible to fully profit from the experimental precision. In contrast, the top quark pair production cross-section near threshold can be calculated with a high degree of precision using theoretically well-defined top mass definitions, such as the 1S mass of Hoang et al.\ \cite{Hoang:1999zc}, which then can be transformed into other commonly used mass schemes. A scan of the $t\bar{t}$ threshold thus provides the possibility for a precise mass measurement with well-controlled theory uncertainties, likely making it the ultimate measurement of the top mass at a linear collider. The measurement of the top mass both at and above threshold with small experimental uncertainties will also provide constraints on the non-perturbative effects in the invariant mass interpretation, providing an important added value beyond the determination of the top quark mass itself.

In this paper, we study the capabilities of a linear $e^+e^-$ collider based on CLIC technology \cite{CLICCDR_vol1} for precise measurements of the top quark mass using both experimental approaches. The studies are performed using simulations based on GEANT4 \cite{Agostinelli:2002hh} of signal and relevant background processes  including machine-related backgrounds in a detailed detector model with realistic modeling of the detector response and with event reconstruction. In Section \ref{sec:ExperimentalConditions}, the experimental conditions and the detector concepts at CLIC are outlined, followed by a summary of the event generation, simulation and reconstruction in Section \ref{sec:Generation} and a description of the techniques used for the identification and reconstruction of top quark events in Section \ref{sec:Reconstruction}. The results for the top mass obtained by direct reconstruction of the invariant mass at $\sqrt{s} = 500$ GeV are given in Section \ref{sec:AboveThreshold} while the threshold scan at energies around 350 GeV is described in Section \ref{sec:Threshold}. The latter also includes a study of the impact of the CLIC luminosity spectrum on the threshold scan compared to that of the ILC, which has a somewhat more peaked spectrum, performed by repeating this analysis using the ILC spectrum.

\section{Experimental conditions and detectors at CLIC}
\label{sec:ExperimentalConditions}

The Compact Linear Collider CLIC is a collider concept based on normal conducting accelerating cavities and two-beam acceleration \cite{CLICCDR_vol1}. It is designed to provide collision energies up to 3 TeV. The project is foreseen to be implemented in several stages \cite{Lebrun:2012hj} to provide optimal luminosity conditions at lower energies. Here, we consider a possible first stage of CLIC with a maximum design energy of 500 GeV, to be operated both at the $t\bar{t}$ threshold around 350 GeV and at 500 GeV. 

The experimental conditions at CLIC are mainly influenced by three factors: the luminosity spectrum, background from two-photon processes and the bunch crossing frequen\-cy of 2 GHz. The small beam spot size required to achieve a luminosity of $2.3 \times 10^{34}$ cm$^{-2}$s$^{-1}$ leads to substantial energy losses due to beamstrahlung, resulting in a luminosity spectrum with 61\% of the luminosity in the top 1\% of the energy at 500 GeV. When operated at 350 GeV, the  luminosity is $1.1 \times 10^{34}$ cm$^{-2}$s$^{-1}$, with 77\% of that in the top 1\%. Radiative effects, together with the high beam energy, result in substantial cross sections for two-photon processes. Of particular relevance are $\gamma\gamma \to$ hadrons processes, which deposit additional energy in the full detector acceptance. At 500 GeV,  such events have an average total energy of 13.3 GeV, 3.4 GeV of which is deposited in the calorimeter system. About 0.3 events on average are expected per bunch crossing. Together with the short bunch-to-bunch spacing of 0.5 ns, this leads to a pile-up of background from many bunch crossings over the 177 ns long bunch trains for typical detector integration times of the order of a few to several tens of nanoseconds. At an energy of 350 GeV, the rate of $\gamma\gamma \to$ hadrons is reduced to 0.05 events per bunch crossing.

The detector concepts for CLIC \cite{Linssen:2012hp} are based on the two ILC concepts ILD \cite{Abe:2010aa} and SiD \cite{Aihara:2009ad}, since the performance requirements for ILC and CLIC are similar. For the CLIC detectors, design modifications motivated by the higher energy of up to 3 TeV and by the more challenging experimental conditions such as a tighter bunch spacing of 0.5 ns and a higher rate of incoherent pairs are implemented. The detector concepts provide highly efficient tracking with excellent momentum resolution in a  solenoidal field, precise secondary vertex reconstruction and highly segmented calorimeters optimized for jet reconstruction using particle flow algorithms. To provide the depth of the calorimeters necessary for multi-TeV operation, the barrel hadronic calo\-rimeter uses tungsten absorbers. The inner radius of the vertex detectors is increased compared to the ILC to account for the larger beam crossing angle and the higher rate of incoherent $e^+e^-$ pairs produced in the collision. While the calorimeters are expected not to change for the different energy stages of CLIC, the vertex detector used at 500 GeV and below can be closer to the interaction point than the one used at 3 TeV. 

For the present analyses, the 500 GeV version of the CLIC\_ILD \cite{Linssen:2012hp} detector is used. This detector concept uses a low-mass pixel vertex detector with an innermost radius of 25 mm, intermediate silicon strip tracking, a TPC as main tracker, a silicon-tungsten electromagnetic calorimeter and a hadron calorimeter system with scintillator tiles read out by silicon photomultipliers with tungsten absorbers in the barrel and steel absorbers in the endcap region. The full detector system is located within a 4 T solenoid, with additional muon tracking in the instrumented return yoke. A detailed model of the detector has been implemented in GEANT4, and the full reconstruction software including particle flow reconstruction has been used in the present analyses. 

\section{Event generation, simulation and reconstruction}
\label{sec:Generation}

The signal process studied here is top quark pair production, $e^+e^- \to t\bar{t}$, which, at a 500 GeV CLIC collider, has a cross section of 530 fb. The top quark decays almost exclusively into a $W$ boson and a $b$-quark. The signal events can thus be grouped into different classes, according to the decay of the $W$ bosons. These are 
\begin{itemize}
\item the {\em fully-hadronic} channel $e^+e^- \rightarrow t\bar{t} \rightarrow q\bar{q}b\,q\bar{q}\bar{b}$, with both $W$s decaying into quarks with a branching ratio of 46\%
\item the {\em semi-leptonic} channel $e^+e^- \rightarrow t\bar{t} \rightarrow q\bar{q}b\, l\nu b$ , with one $W$ decaying into quarks, the other into a lepton and a neutrino with a branching ratio of 15\% for each of the three lepton flavors
\item the {\em fully-leptonic} channel $e^+e^- \rightarrow t\bar{t} \rightarrow l\nu b l\nu \bar{b}$, with both $W$s each decaying into a lepton and a neutrino, with a branching ratio of 9\%. 
\end{itemize}
In the leptonic channels, the decay into a $\tau$ and a neutrino is a special case, since the $\tau$ decays almost instantly into either a lepton and two neutrinos or into one or more hadrons and a neutrino, giving rise to additional missing energy in the final state, and potential confusion with hadronic decay modes. 

In the analyses discussed below, only fully-hadronic and semi-leptonic events, excluding $\tau$ final states, are selected, since these provide the best mass resolution and the clearest identification of top quark events. These correspond to 76\% of all top quark pair decays. However, to account for imperfect event classification, all possible decay modes of the $t\bar{t}$ pair are generated according to their branching fractions and included in the signal sample. The top mass and width are fixed for the signal event sample generation to $m_{\rm{top}}$ = 174.0 GeV and $\Gamma_{\rm{top}}$ = 1.37 GeV. 

\begin{table}
\centering
\begin{tabular}{c|c|c|c}
\hline
type & final & $\sigma$ & $\sigma$\\
& state & 500 GeV & 352 GeV\\
\hline
\hline
Signal ($m_{\rm{top}}$ = 174 GeV)& $t \bar t$	& 530 fb& 450 fb\\
\hline
Background& $W W$	& 7.1 pb & 11.5 pb	\\
Background& $Z Z$	&  410 fb & 865 fb \\
Background& $q \bar q$	&  2.6 pb & 25.2 pb\\
Background& $W W Z$	&  40 fb & 10 fb\\
\end{tabular} 
\caption{Signal and considered physics background processes, with their approximate cross section calculated for CLIC at 500 GeV and at 352 GeV.}
\label{tab:channel-production}
\end{table}

In addition to the signal, background processes with similar event topologies have to be considered. These are four and six fermion final states from gauge boson production. Because of its high cross section, the two-fermion quark pair production, which can produce a signal-like topology in badly reconstructed events, is also considered. Table \ref{tab:channel-production} lists the studied processes, with cross sections at a CLIC machine operated at 500 GeV and 352 GeV.  The background samples are generated for an integrated luminosity of 100 fb$^{-1}$, with additional events produced for the signal process to provide high-statistics data sets for signal-only studies to determine the detector resolution function. In addition, the processes  $e^+e^- \to q \bar q e^+ e^-$ and $e^+e^- \to q \bar q e \nu$, which are dominated by t-channel single boson production, are investigated. It is shown that the non-di-boson contributions are rejected completely in the analysis. Since the di-boson contributions are accounted for in the $e^+e^- \to WW$ and  $e^+e^- \to ZZ$ modes, these additional final states are not considered separately. 

In addition to the main production of signal and background processes, a signal sample with a different top quark mass and width ($m_{\rm{top}}$ = 175 GeV, $\Gamma_{\rm{top}}$ = 1.5 GeV) is generated to study possible systematic effects in the results depending on the input parameters used to derive the detector resolution function. 

PYTHIA \cite{Sjostrand:2006za} is used to generate the signal process $e^+e^-$ $\to t\bar{t}$ as well as the two background processes $e^+e^- \to WW$ and  $e^+e^- \to ZZ$. PYTHIA correctly accounts for a non-zero width of explicitly defined intermediate and final states. The processes $e^+e^- \to q \bar q$, $e^+e^- \to q\bar{q} e^+e^-$ and $e^+e^- \to q\bar{q} e \nu$ are generated using WHIZARD \cite{Kilian:2007gr}, which was used as the default event generator for the benchmark studies for the CLIC Conceptual Design Report. WHIZARD is also used for the background process  \mbox{$e^+e^- \to WWZ$}. For simplicity, these events are generated with zero width for the bosons, which allows to specify the tri-boson final state explicitly prior to the decay into six fermions. 

For all generated events, the luminosity spectrum of the 500 GeV CLIC machine is considered, and beam-induced background in the form of  $\gamma \gamma \to$ hadrons events is added. More information on the beam induced backgrounds can be found in \cite{Linssen:2012hp}. The events are simulated with GEANT4  using the CLIC\_ILD detector model  defined in Mokka \cite{MoradeFreitas:2002kj}. 300 bunch crossings of $\gamma \gamma \to$ hadrons events are overlaid with the signal event at the digitization stage. The tracking and particle flow event reconstruction \cite{Marshall:2012ry} is then performed on the combined event comprising signal and beam-induced background.  
To reduce the impact of this background, time-stamping in the detector subsystems is used to assign energy deposits to individual bunch crossings. Since the background particles are predominantly at low $p_{\rm{T}}$ in the forward and backward regions of the detector, transverse momentum requirements dependent on the polar angle are used to further reduce the effect of the background. The applied cuts are documented in \cite{Linssen:2012hp, Marshall:2012ry}. The data analysis is then performed on reconstructed particles (Particle Flow Objects - PFOs) after the application of these cuts.

\section{Top quark reconstruction at CLIC}
\label{sec:Reconstruction}

The identification of top quark pair production events and the rejection of non-$t\bar{t}$ background proceeds in several steps. The analysis strategy outlined in the following is used both above and at threshold. The event selection criteria differ for the two mass measurement methods, since for the direct reconstruction of the mass the quality of the selected events is of key importance, while for the cross section measurement in a threshold scan the emphasis is on maximising the signal significance.  

As a first step, all events are classified according to the number of isolated leptons (electrons and muons) in order to identify the top pair decay modes $t\bar{t} \rightarrow q\bar{q}b\,q\bar{q}\bar{b}$ and  $t\bar{t} \rightarrow q\bar{q}b\, l\nu b$ which are used further in the analysis.  The events are classified as fully-hadronic (no isolated lepton found), semi-leptonic (exactly one isolated lepton found) or fully-leptonic (at least two isolated leptons found). Events classified as fully-leptonic are rejected, while the other two classes are clustered into four or six jets, according to event class. Subsequently, a flavor tagging algorithm is used to identify the two jets most likely originating from $b$-quarks.  For the fully-hadronic channel the correct combination of the four non-$b$-jets into $W$ bosons has to be found among the three possible combinations, while for the semi-leptonic case the assignment of light jets and leptons to $W$ candidates is unique. The pairing of $W$ candidates and $b$-jets into the two top candidates is done with a kinematic fit that exploits energy and momentum constraints to improve the top mass measurement. This step also provides strong rejection of non-$t\bar{t}$ background since such events tend to lead to a failure of the fit by not satisfying the imposed constraints. Additional background rejection is achieved with a binned likelihood technique, after which a highly pure sample of top quark pair events is available for the reconstruction of the invariant mass or the measurement of the cross section. In the following, each of these steps is described. 

\subsection{Lepton finding}

The classification into fully-hadronic, semi-leptonic and ful\-ly-leptonic events is based on the identification of isolated leptons using a lepton finder as a first step of the analysis. It is optimized to identify charged leptons ($e^\pm$ or $\mu^\pm$) originating from the decay of $W$ bosons. Since these leptons are typically highly energetic, and, in contrast to leptons originating from hadronic decays in quark jets, well separated from other activity in the event, isolation and energy are used as selection criteria. A minimum lepton energy of 10 GeV is required. If there is no other charged particle with an energy larger than 2.5 GeV within a cone with an opening angle of $10^\circ$ around the lepton track, it is considered to be isolated. The cut values were optimized in a parameter scan with $t\bar{t}$ events using generator level simulation information.

The residual $\gamma \gamma \to$ hadrons background leads to a slight deterioration of the quality of the event classification, reducing efficiencies by 1\% to 2\% compared to the case without additional background. Overall, 96\% of the all-hadronic events and 91\% of the semi-leptonic events (ignoring final states with $\tau$ leptons) are correctly classified. At this stage, 56\% of all input events are classified as fully-hadronic and 33\% as semi-leptonic candidates. The remaining 11\% are rejected. Out of the accepted events, in particular the fully-hadronic, but also the semi-leptonic samples receive contributions from events with $\tau$ final states which are reduced in further steps of the analysis.

\subsection{Jet clustering}

Jet clustering of the events is performed in exclusive mode, meaning that events are clustered into a fixed number of jets. According to the classification of the lepton finder, events are clustered either into 6 jets (fully-hadronic) or into 4 jets (semi-leptonic).  In the latter case, the identified isolated lepton is excluded from jet finding. 

Jet finding is performed with the $k_{\rm{t}}$ algorithm \cite{Ellis:1993tq} implemented in the FastJet package \cite{Cacciari:2005hq, Cacciari:2011ma}. This algorithm uses the pseudorapidity $\eta$ and the azimuthal angle $\phi$ as the basis to define the two-particle distance. This metric results in an increase of the two-particle distance in the forward region and makes the algorithm robust against $\gamma \gamma \rightarrow$ hadrons background since these mainly forward-going particles tend to be clustered into the beam jets. These are implemented in the algorithm to account for beam remnants in the case of hadron collisions and are not considered further in the analysis. In the present study, the  $k_{\rm{t}}$ jet algorithm is used with a jet size parameter $R$ of 1.3, which is selected as the best trade-off between the requirements to limit the loss of signal particles on the one hand and to limit the inclusion of background on the other.

\subsection{Flavor tagging}

Efficient $b$-tagging is essential for the identification of $t \bar t \rightarrow (bq\bar q) (\bar b q\bar q)$ and $t \bar t \rightarrow (bq\bar q)(\bar b l \nu_l)$ events compared to multi-fermion background, and is also crucial for the correct assignment of jets to top candidates for signal events. In the analysis, flavor tagging is performed using the {\it LCFI Flavour Tagging} \cite{Bailey:2009ui} package. This algorithm is based on a neural network which provides $b$ and $c$-jet probabilities (``$b$-tag'') for each jet in the event, depending on a number of input variables such as reconstructed secondary vertices, particle momenta and impact parameters. The training of the neural network used for the tagging is performed with a fully-hadronic sample of $t \bar t$ events. These events are generated and reconstructed without beam spectrum, initial state radiation and top width, but do contain all other generation, simulation and reconstruction details, such as overlaid $\gamma \gamma \to$ hadrons background.

\subsection{Jet pairing}

\begin{figure}
\centering
  \includegraphics[width=0.99\columnwidth]{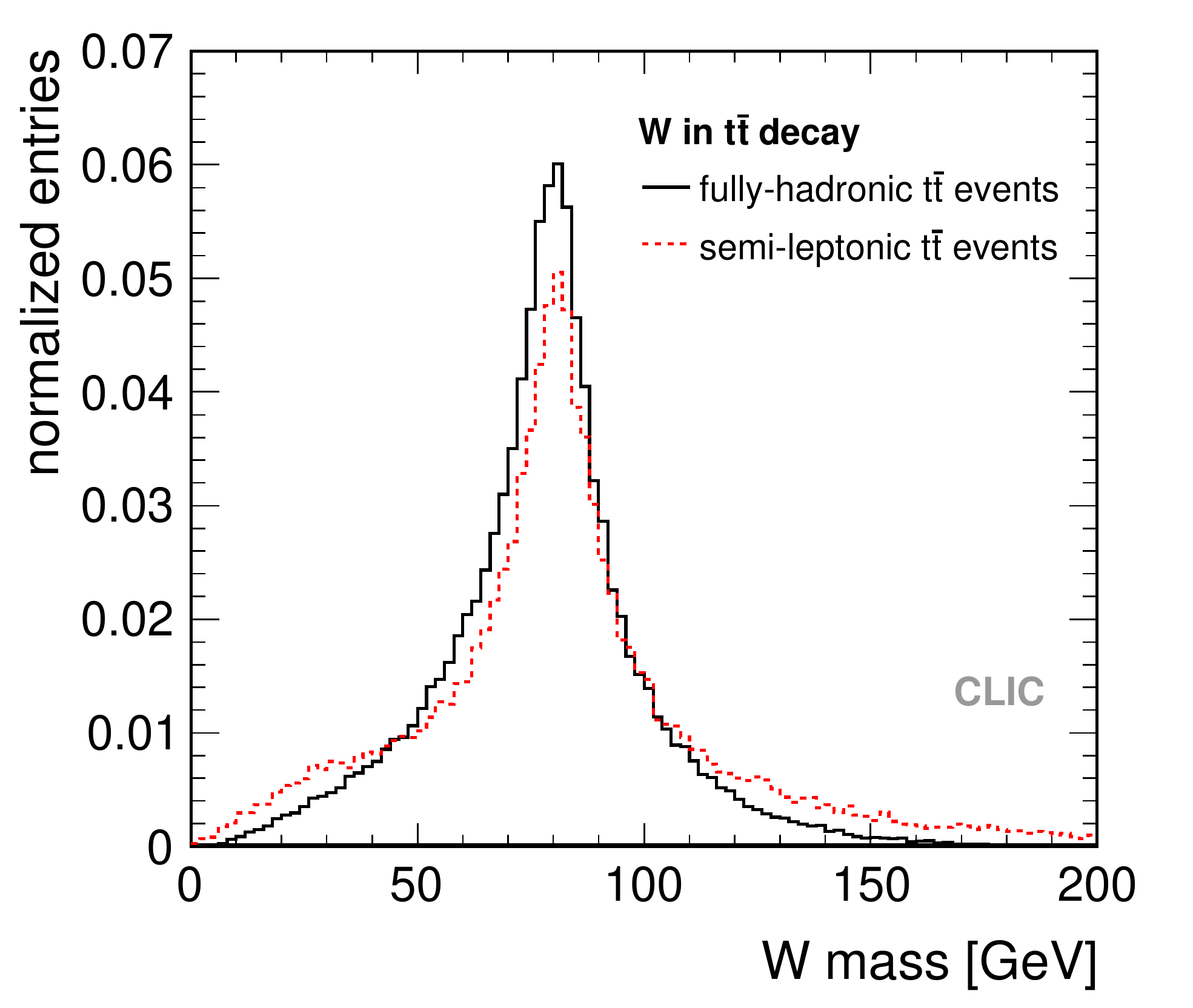}
	\caption{Distribution of the reconstructed mass of the $W$ bosons for a signal-only sample after the classification into fully-hadronic and semi-leptonic candidates before further event selection for collisions at 500 GeV. The distributions are normalized to allow a shape comparison.}
   \label{fig:WMasses}
 \end{figure}

For both analysis branches (6-jet fully-hadronic and 4-jet semi-leptonic candidates), the two jets with the highest $b$-tag values are classified as jets created by a $b$-quark ($b$-jets). All other jets in the event are classified as light jets (created by $u$, $d$, $s$ or $c$-quarks) originating from the $W$ boson decay.
 
 In the case of the 4-jet sample all decay products of the $t \bar t$ pair are found at this stage of the analysis: Two $b$-jets, two light jets forming one $W$ boson, one charged lepton and a neutrino forming the other $W$ boson. As the neutrino is not measured, its momentum is taken to be the total missing momentum in the event. Since the neutrino absorbs all energy reconstruction uncertainties, and since the missing energy measurement also includes contributions from the luminosity spectrum, the leptonic $W$ boson mass is less well constrained than the hadronic $W$ boson mass.  In the 6-jet case, the correct pairing of light jets into $W$ bosons has to be identified among the three possible permutations of light jet pairs. For each permutation the invariant mass of both jet pairs is calculated and compared with the true $W$ boson mass. The permutation with the minimal value of 
  \begin{eqnarray}
 \label{eq:jet-combi}
 v = | m_{ij} - m_W |  + | m_{kl} - m_W |, 
 \label{eq:WMass}
 \end{eqnarray}
 where $m_W$ = 80.4 GeV and $m_{ij}$ and $m_{kl}$ are the invariant masses of two distinct jet pairs, is chosen as the correct combination. 
 
 Figure \ref{fig:WMasses} shows the normalized distribution of the reconstructed invariant mass for all $W$ candidates in a signal only sample. For the fully-hadronic analysis branch, only the $W$ candidates chosen according to Equation \ref{eq:WMass} are shown. The distribution for the semi-leptonic candidates has more pronounced tails due to the aforementioned additional uncertainties entering for the neutrino in the leptonically decaying $W$.

\subsection{Kinematic fitting}
\label{sec:KinFit}

After the identification of $b$-jets and the pairing of light jets and leptons into $W$ bosons, the next step of the analysis is the correct grouping of $W$ boson candidates and $b$-jets into top quarks. This assignment is performed using a kinematic fit. Out of the two possible combinations, the one with the higher probability of the kinematic fit result is chosen. 

The kinematic fit \cite{Beckmann:2010ib} uses constraints based on the assumption of a $t \bar t$ event to improve the precision of the event parameters of interest. The parameters of interest are fitted using a least squares technique and the physical constraints are incorporated into the fit using Lagrange multipliers. In the present analysis, all constraints are imposed as hard constraints, meaning that they have to be fulfilled exactly. The input to the kinematic fit in the case of the 6-jet sample are the four-momenta of the light jets, already paired into $W$ bosons, and the four-momenta of the $b$-jets. In the case of the 4-jet sample, the input values are the four-momenta of the light jets and $b$-jets as well as of the isolated lepton. In the latter case,  the unmeasured neutrino is also represented in the fit, with starting values set to the measured missing energy and momentum in the event. The constraints imposed are full three-dimensional energy and momentum conservation taking into account the crossing angle of the beams, the correct reconstruction of the $W$ boson masses to the nominal mass of $m_W$ = 80.4 GeV, and an equal mass of the two top candidates.  In order to fulfill these constraints, the kinematic fit varies the reconstructed energies and angles of the jets and leptons. These variations are constrained by the experimental jet energy and angular resolution and the lepton momentum and angular resolution for jets and leptons, respectively.

The kinematic fit fails if it is unable to satisfy the constraints outlined above, which is typically due to mistakes in the event reconstruction. Since badly reconstructed events will not yield a correct mass value, these events are rejected in the analysis. The failures are due to several effects such as the wrong classification into the semi-leptonic and fully-hadronic event branch, imperfect jet clustering, $W$ boson reconstruction from the wrong jet combination, too large remaining $\gamma \gamma \to$ hadrons background and the influence of the luminosity spectrum. 

It is observed that some of the fit failures are due to the wrong identification of one of the $b$-jets. This is particularly likely in the case of a $W$ decaying into a charm quark and another light quark. Thus, to improve the number of successful fits and to account for possible wrong flavor tagging, the jet pairing into $W$ candidates and the kinematic fit are repeated for unsuccessful kinematic fits after exchanging the $b$-jet with the lower $b$-tag value and the light jet with the highest $b$-tag value. This procedure increases the number of successful kinematic fits by approximately 20\%.

\begin{figure}
\centering
  \includegraphics[width=0.99\columnwidth]{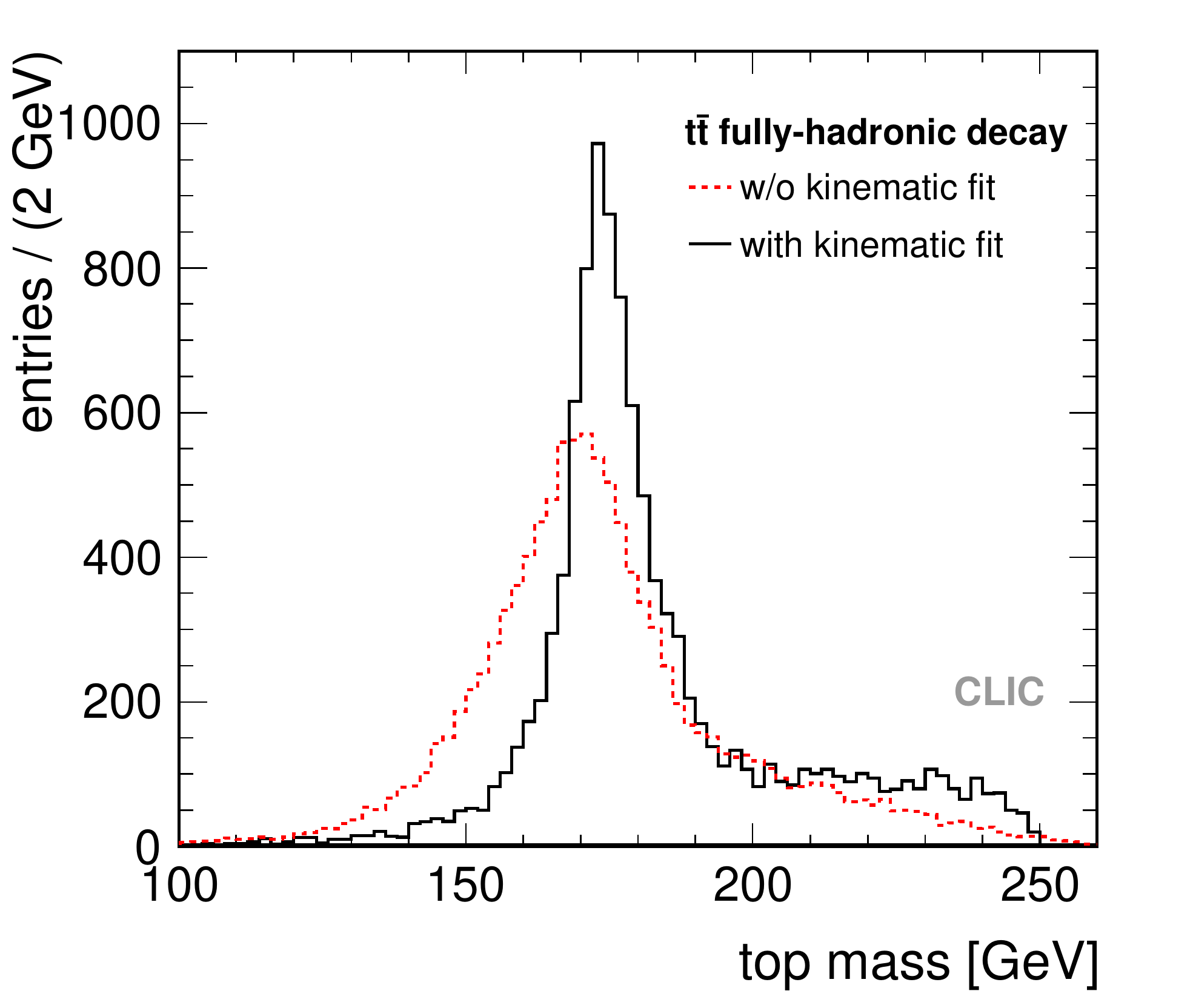}
	\caption{Reconstructed top mass for signal-only events selected in the fully hadronic branch with and without kinematic fit for collisions at 500 GeV. For both distributions only events with a successful kinematic fit are shown.}
   \label{fig:KinFitFH}
 \end{figure}

Overall, 37\% of the signal events classified as fully-ha\-dron\-ic candidates and 47\% of the signal events classified as semi-leptonic candidates pass the kinematic fit successfully at 500 GeV.  This failure rate reflects the optimization of the analysis towards cleanly reconstructed events to provide the best possible mass measurement. Figure \ref{fig:KinFitFH} illustrates the influence of the kinematic fit on the reconstructed top quark mass by comparing the mass distributions before and after the kinematic fit for fully-hadronic events with successful fits.

At threshold, fit success rates of  95\% are obtained for the fully-hadronic and the semi-leptonic analysis branch. This higher success rate is due to more constrained kinematics directly at the production threshold and reduced two-photon backgrounds as well as to the acceptance of events with converged kinematic fits regardless of the fit probability. These reduced requirements are possible since at threshold the highest possible significance and not the cleanest possible reconstruction is the goal of the analysis.

By imposing requirements on the kinematic structure of the events, the kinematic fit also serves as a powerful tool for rejecting non-$t\bar{t}$ background. More than 97\% of the $q\bar{q}$ background is rejected by the fit in both branches, and more than 94\% of the di-boson pair production events in the fully-hadronic analysis branch. In the semi-leptonic branch, $ZZ$ events are rejected with an efficiency of 94\%, $WW$ events with 86\%. The $WWZ$ background, which can result in final states identical to those of top pair events and has similar kinematics overall, is only rejected at the level of 13\% for fully-hadronic and 25\% for semi-leptonic events.

\subsection{Additional background rejection}

Additional signal and background discrimination is provided by a binned likelihood technique  \cite{Ackerstaff:1997cza} which combines several discriminating observables into one likelihood variable. For the two event classes $j$, ``signal'' and ``background'', probability density functions $f^j(x_i)$ for each discriminating variable $x_i$ are provided as input to the likelihood algorithm.  The probability $p^j(x_i)$ for a given event to belong to event class $j$ for a given value of the discriminating variable $x_i$ is given by  
\begin{equation}
\label{eq:prob-func}
p^j(x_i) = \frac{f^j(x_i)}{\sum_k	 f^k(x_i)},
\end{equation}  
where $k$ runs over all event classes.

The final likelihood for an event belonging to the signal event class S, combining the probabilities of the individual discrimination variables, is given by  
\begin{equation}
\label{eq:likelihood}
L_S = \frac{\prod_i p^S(x_i)}{\sum_k	\prod_i p^k(x_i)},
\end{equation} 
with $i$ running over all discrimination variables and $k$ over all event classes. 

The variables entering the likelihood are the two highest $b$-tags, the sphericity and the overall number of particles in the event, the masses of the $W$ bosons, the difference of the two top masses without kinematic fit, and the d$_{\mathrm{cut}}$ variable of the jet clustering algorithm. The d$_{\mathrm{cut}}$ variable provides a measure of how natural it is to cluster a given event into the defined number of jets rather than fewer jets. Since $t\bar{t}$ events are characterized by a higher jet multiplicity than most of the background channels, this variable provides a good separation of signal and background. 

For signal events passing the kinematic fit, an efficiency of 93.5\% (94.4\%) is achieved for the fully-hadronic (semi-leptonic) analysis branch. At the same time, 98.0\% (96.7\%) of the background events remaining after the kinematic fit are removed by this additional background rejection step in the fully-hadronic (semi-leptonic) event sample. Together with the kinematic fit,  99.8\% (99.7\%) of all background events are rejected. 

\section{Top mass measurement above threshold}
\label{sec:AboveThreshold}

\begin{figure*}
\centering
  \includegraphics[width=0.99\columnwidth]{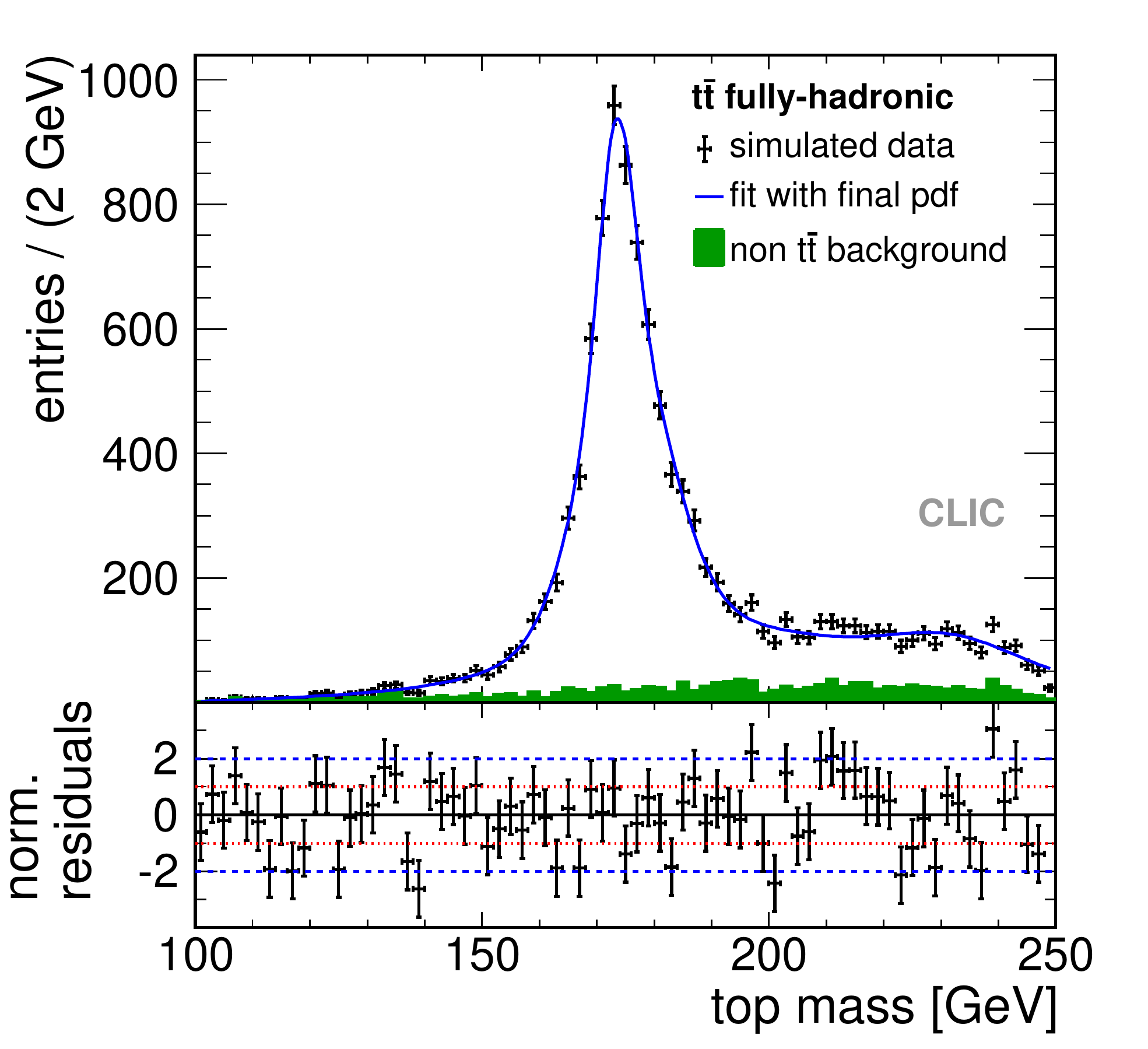}
   \includegraphics[width=0.99\columnwidth]{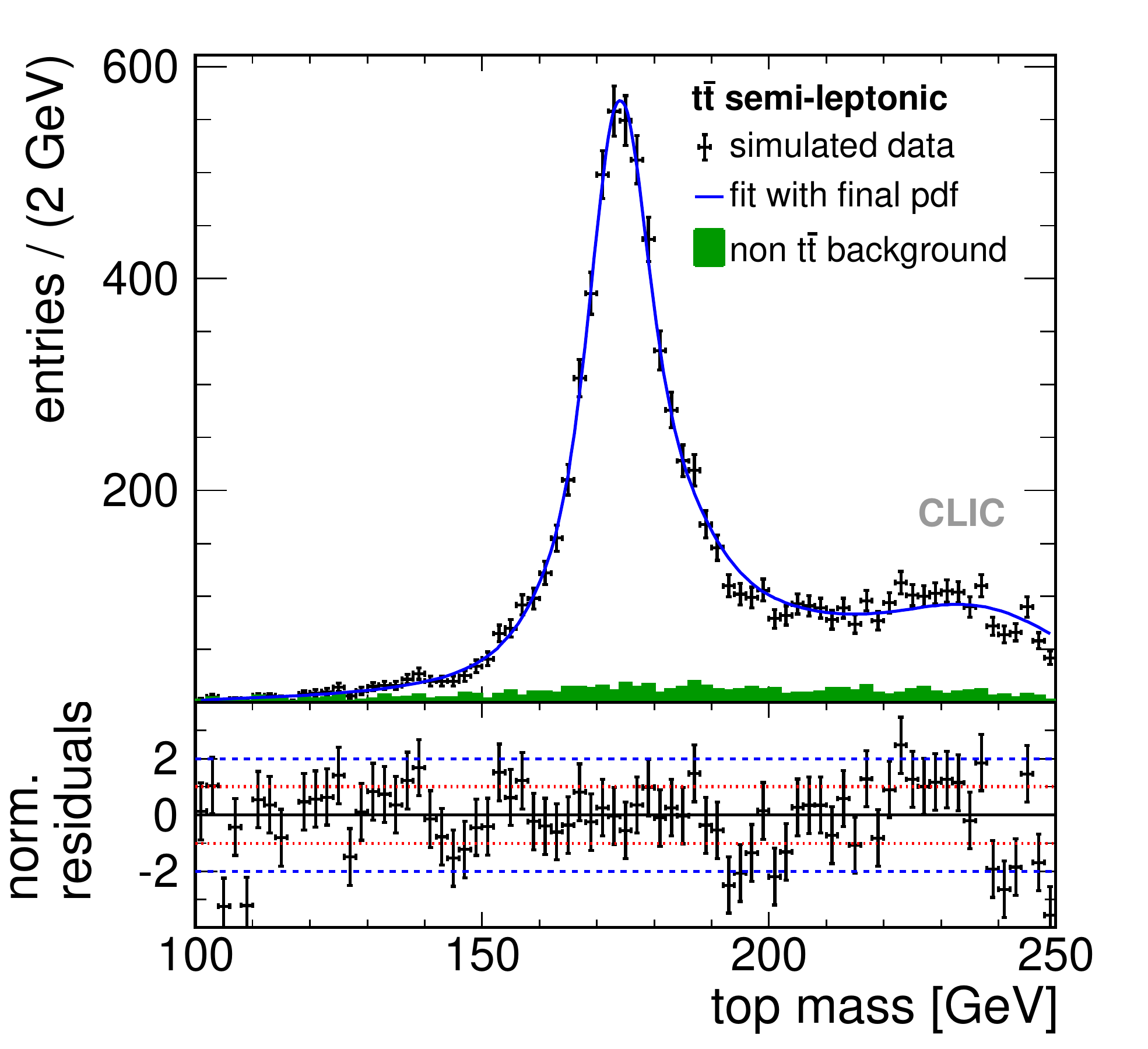}
	\caption{Distribution of reconstructed top mass for events classified as fully-hadronic (left) and semi-leptonic (right). The data points include signal and background for an integrated luminosity of 100 fb$^{-1}$. The pure background contribution contained in the global distribution is shown by the green solid histogram. The top mass is determined with an unbinned likelihood fit of this distribution, which is shown by the solid line.}
   \label{fig:TopMass}
 \end{figure*}

At energies substantially above the production threshold, in this paper studied at 500 GeV, the top quark mass is measured from the invariant mass of the reconstructed decay products. 

\subsection{Determination of top quark mass and width}

In this analysis, the distribution of the reconstructed invariant mass for signal and background events after kinematic fit and background rejection is considered. Since the kinematic fit imposes the constraint of equal masses of the top and the anti-top in the $t\bar{t}$ pair, each event provides one unique measurement of the mass and thus enters only once in the distribution.

Figure \ref{fig:TopMass} shows the reconstructed top mass for signal and background after kinematic fitting and background rejection for an integrated luminosity of 100 fb$^{-1}$, recorded at \mbox{500 GeV}. 
From this distribution, the top mass is extracted using an unbinned likelihood fit. The fit function consists of three components, the first of which accounts for physics background, while a convolution of the other two describes the resolution and the signal itself. Prior to the final fit, a parametrization of the background distribution is obtained from a fit to a background-only sample and the influence of the detector resolution is determined with a statistically independent signal-only sample corresponding to twice the integrated luminosity considered in the analysis.

The background is parametrized by a  threshold function
\begin{eqnarray}
  bgd = (x - a )^b
\end{eqnarray}
in which the threshold is fixed to $a$ = 100 GeV and $b$ is left as a free parameter which is determined with a fit to a background-only sample.

For the signal fit, the following function is used:
\begin{eqnarray}
   signal &=& f \cdot \textrm{BreitWigner}(m_{\rm{bw}}, \Gamma_{\rm{bw}})  \otimes \ res \\
 		      & & + (1 -f ) \cdot \ tail \nonumber
\end{eqnarray}
This fit consists of two main components, a signal part described by a Breit-Wigner convolved with a  resolution function {\it res} implemented as the sum of three Gaussians, and a background part labeled {\it tail} to account for the high-mass tail observed in the signal at masses around 230 GeV. This tail is due to kinematic reflections in events with incorrect assignments of jets and/or leptons to top candidates. The mean value of the Breit-Wigner is given by $m_{\rm{bw}}$, and the corresponding width by $\Gamma_{\rm{bw}}$. It is found that the tail is well modelled by a simple Gaussian. The relative fraction of the two components is described by the factor $f$. 

The resolution component of the signal part is described by
\begin{eqnarray}
 res &=& f1 \cdot \textrm{Gauss1}( x, m1, s1 ) + \\
				& &      f2 \cdot \textrm{Gauss2}( x, m2, s2 ) + \nonumber \\ 
	                             & & (1 - f1-f2) \cdot \textrm{Gauss3}( x, m3, s3 ), \nonumber
\label{eq:resolution-function}
\end{eqnarray}
where the notation Gauss(x, mean, sigma) is used and $f$1 and $f$2 are the fractions of the Gaussians in the sum. This function does not only represent the detector resolution, but also accounts for systematic effects introduced by the analysis chain and by the pick-up of $\gamma\gamma \rightarrow$ hadrons background. The resolution function is determined with a fit to a separate signal-only event sample for which the mean and the width of the Breit-Wigner component are fixed to the generator values of $m_{\rm{bw}}$ = 174 GeV and $\Gamma_{\rm{bw}}$ = 1.37 GeV. All parameters of the Gaussian sum are left free in this fit. 

The full probability density function of the final fit  for the top mass distribution containing signal and background is given by the sum of the signal and background functions, 
 \begin{eqnarray}
  pdf_{\rm{fit}} = y_{\rm{sig}} \cdot signal + y_{\rm{bgd}} \cdot bgd, 
 \end{eqnarray}
where $y_{\rm{sig}}$  and  $y_{\rm{bgd}}$ describe the signal and the background yield, respectively. In the final fit to the overall top mass distribution including signal and background contributions, fixed values for the Gaussians of $signal$ and $bgd$ are used, leaving only $m_{\rm{bw}}, \Gamma_{\rm{bw}}$,  $y_{\rm{sig}}$ and $y_{\rm{bgd}}$ as free parameters. The fit, shown by the lines in Figure \ref{fig:TopMass}, is performed independently for the fully-hadronic and for the semi-leptonic events.

\begin{table}
\centering
\begin{tabular}{l|c|c|c|c}
\hline
channel & $m_{\rm{top}}$& $\Delta m_{\rm{top}}$& $\Gamma_{\rm{top}}$ & $\Delta\Gamma_{\rm{top}}$ \\
\hline
fully-hadronic & 174.049 & 0.099 & 1.47 & 0.27\\
semi-leptonic & 174.293 & 0.137 & 1.70 & 0.40\\
combined & 174.133 & 0.080 & 1.55 & 0.22\\
\hline
\end{tabular}
\caption{Summary of the top mass measurement at 500 GeV for an integrated luminosity of 100 fb$^{-1}$. All numbers are given in units of GeV, Errors are statistical only.}
\label{tab:AboveThresholdResults}
\end{table}

The results of the fit are summarized in Table \ref{tab:AboveThresholdResults}. The determined masses and widths are in good agreement with the generator values, with a $2 \sigma$ deviation observed for the reconstructed mass in semi-leptonic events. Combining the results of both event classes, a statistical uncertainty of 80 MeV for the top quark mass is obtained for an integrated luminosity of 100 fb$^{-1}$. 

The uncertainty on the width is considerably larger since the width of the reconstructed mass distribution is  dominated by the resolution function, which incorporates detector and reconstruction effects as well as the influence of the collider luminosity spectrum. Still, a  combined uncertainty of 220 MeV is reached with the present fit function.

\subsection{Systematic studies}

While a  full study of possible systematic uncertainties is beyond the scope of the current work, two potentially sizable sources of systematic uncertainties have been investigated: the influence of the assumed mass and width in the adjustment of the mass fit procedure, which accounts for the detector resolution and reconstruction effects, and the jet energy scale. As discussed in Section \ref{sec:Intro}, there are sizeable uncertainties in the theoretical interpretation of the measured invariant mass value, which potentially surpass the statistical and systematic experimental uncertainties discussed here. All uncertainties discussed for the top mass measurement at 500 GeV apply to the top mass parameter as implemented in the event generator.

A possible bias due to the adjustment of the fit procedure used to determine the mass from the invariant mass distribution is studied by using a data sample with a top quark mass of 175 GeV and a width of 1.5 GeV, compared to 174 GeV and 1.37 GeV used in the signal sample in the analysis, for the determination of the fit parameters. The results obtained with these parameters deviate from the original results by less than the statistical uncertainty for both fully-hadronic and semi-leptonic events. Since the statistical uncertainty of the detector resolution parameters themselves is comparable to the overall statistical uncertainties due to comparable integrated luminosities in the training and in the signal samples, there is no indication for a bias introduced by the {\it a priory} uncertainty of the top quark mass. While small biases can not be excluded due to the limited statistics, they are below the statistical uncertainties of the analysis. 

The jet energy scale is one of the most important sources of systematic uncertainties in top quark mass measurements to date. To investigate this effect on the present analysis, three separate studies have been performed: the energy of all jets was shifted by 2\%, the energy of light jets only was shifted by 2\%, and the energy of b jets only was shifted by 2\%. The kinematic fit, which imposes constraints on the reconstructed $W$ mass, limits the influence of a shift in the light jet scale to less than 100 MeV. Correlated shifts of all jets result in a bias of 200 MeV, while a shift of b-jets only, which are not well constrained by the kinematic fit, results in a bias of 350 MeV. The influence of shifts in the light jet energy scale can be further reduced by an {\it in-situ} calibration based on the reconstructed $W$ mass before the kinematic fit. For an integrated luminosity of 100 fb$^{-1}$, the precision of the mass obtained with a simple fit of the invariant mass distribution is better than 100 MeV, which will allow to constrain the jet energy scale to better than 1\%. This in turn results in a systematic uncertainty of the top mass below 50 MeV. While such a possibility does not exist for b-jets in top quark pair production, it is expected that a similar precision can be reached for b-jets by reconstructing $Z$ decays in di-boson production and radiative return events. The determination of the b-jet energy scale to better than 1\% in such events will bring the total jet energy scale systematics on the top mass measurement at CLIC to a level comparable to the statistical uncertainties of the measurement.

Additional potentially important systematic uncertainties originate from color reconnection \cite{Gustafson:1988fs, Sjostrand:1993hi} between final state partons produced  from the $t$ and $\bar{t}$ decays. Due to the absence of colored beam remnants these effects are expected to be substantially smaller at CLIC than at hadron colliders.  Based on early studies of color reconnection effects in top pair production in $e^+e^-$ collisions \cite{Khoze:1994fu, Khoze:1999up} and based on the size of the uncertainties observed in the $W$ mass determination at LEP2 \cite{Schael:2006mz, Abdallah:2008ad, Achard:2005qy, Abbiendi:2005eq} which have been further constrained with data from all four experiments to 35 MeV for fully hadronic $W$ pair decays \cite{Schael:2013ita}, it is expected that these uncertainties are comparable in size or smaller than the statistical uncertainties of the invariant mass measurement. A dedicated study of these effects is beyond the scope of the present paper.

\section{Top mass measurement in a threshold scan}
\label{sec:Threshold}

In addition to the direct reconstruction of the top quark mass, a linear collider offers the possibility of a threshold scan, which allows to measure the top mass in a theoretically well-defined way, as discussed above.

To determine the top mass in a threshold scan, a measurement of the $t\bar{t}$ production cross section at several points around the threshold is necessary. To identify top pair events with high purity the same analysis procedure as for the invariant mass measurement is used, albeit with relaxed cuts on the quality of the kinematic fit, resulting, together with the changed kinematics and with reduced beam-induced backgrounds, in substantially higher efficiency as detailed in Section \ref{sec:KinFit}. Since the threshold behavior of the cross section also depends on the strong coupling constant \cite{Fadin:1987wz, Fadin:1988fn, Strassler:1990nw}, the top quark mass is extracted simultaneously with $\alpha_s$. Depending on the precision of $\alpha_s$ at the time when such a measurement will be performed, the strong coupling can alternatively be used as external input, as will also be discussed below. Earlier studies \cite{Martinez:2002st} have shown that in addition to the mass and the strong coupling constant, also the top quark width is accessible in a threshold measurement, in particular when using other observables such as the top quark momentum distribution and the forward-backward asymmetry in addition to the cross section. These additional aspects of $t\bar{t}$ threshold measurements are beyond the scope of the present paper. 

For the correct description of the cross section near the production threshold, the inclusion of higher-order QCD contributions is necessary. Since no appropriate event generator is publicly available at present, the study follows the strategy of earlier studies performed for the TESLA collider \cite{Martinez:2002st} by splitting the simulation study into two parts: the determination of the event selection efficiency and background contamination, and the calculation of the top-pair production cross section in the threshold region. In this approach, the signal selection and background rejection is determined using fully simulated top-pair signal events as well as relevant background channels at a nominal center-of-mass energy of 352~GeV, slightly above the production threshold for the selected top mass of 174~GeV. For this, the full simulation, reconstruction and event selection procedure as described in Section \ref{sec:Generation} is followed.  Data points along the threshold curve are then generated by taking the signal cross section determined using NNLO calculations combined with the selection efficiency, adding background events assuming a constant level over the considered energy range of 10~GeV as determined from the full simulations. In the following, more details are given on the individual steps.  

In the analysis, we consider a threshold scan with 10 energy points spaced by 1 GeV from 344 GeV to 353 GeV, with an integrated luminosity of 10 fb$^{-1}$ each, resulting in a total integrated luminosity of 100 fb$^{-1}$.

\subsection{The $t\bar{t}$ threshold at CLIC}

The top-pair signal cross section is determined using full NNLO calculations provided by the code \mbox{{\sc Toppik} \cite{Hoang:1999zc, Hoang:1998xf}}. The top mass input is set to 174~GeV in the 1S mass scheme \cite{Hoang:1999zc}. The strong coupling constant $\alpha_s$ is taken to be 0.118. Since \mbox{{\sc Toppik}} provides the cross section in units of $R$, the ratio of $\sigma(e^+e^-\rightarrow X)$ to  $\sigma(e^+e^-\rightarrow \mu^+\mu^-)$, the appropriate conversion factor of the energy-dependent cross section $e^+e^- \rightarrow \mu^+\mu^-$ is applied in addition. 

\begin{figure}
\centering
\includegraphics[width=0.99\columnwidth]{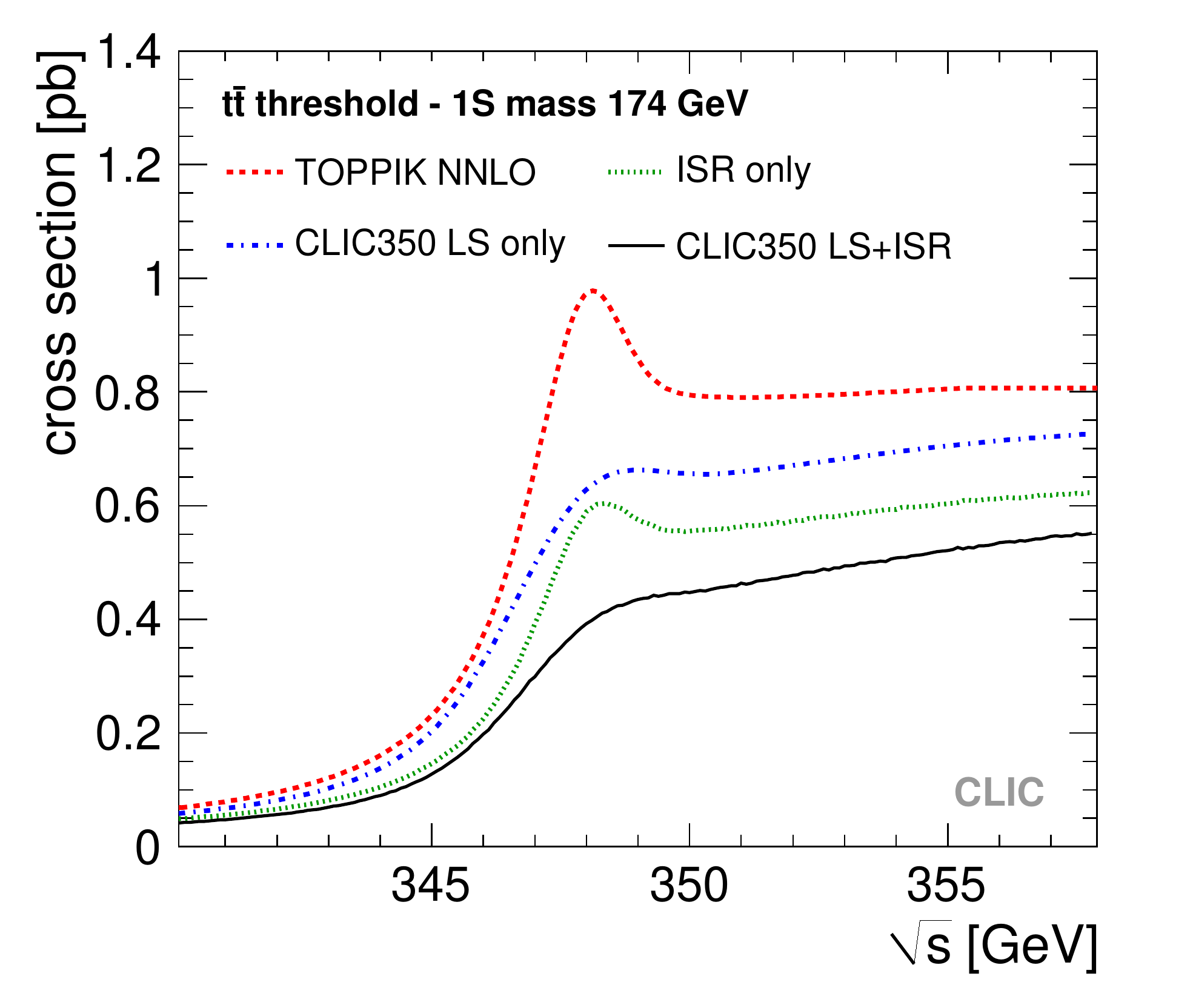}
\caption{Top pair production cross section from theory calculations, with the luminosity spectrum (LS) of CLIC at 350 GeV and ISR as well as for all effects combined.  \label{fig:TopXs}}
\end{figure}

Since this cross section is calculated for the energy at the $e^+e^-$ vertex, additional corrections for initial state radiation (ISR) and for the luminosity spectrum of the collider have to be applied. Initial state radiation is numerically folded into the cross section calculated by \mbox{{\sc Toppik}} following the YFS (Yennie-Frautschi-Suura) solution as given in \cite{Skrzypek:1990qs}. In addition, the luminosity spectrum of CLIC operated at 350 GeV, which is characterized by a main peak containing 77\% of the full luminosity in the top 1\% of the energy and by a long tail to lower energies, is considered. Figure \ref{fig:TopXs} illustrates the influence of these effects  on the cross section. Both ISR and the luminosity spectrum result in a lowering of the cross section since part of the collision events are moved to energies below the threshold. The tail to lower energies, but in particular also the beam energy spread in the main peak of the luminosity spectrum, result in a smearing of the cross section peak at threshold.

\subsection{Generation of data points}

The signal and background efficiencies are determined using fully simulated events as outlined in Section \ref{sec:Generation}. The kinematic fit and the likelihood-based background rejection are used to eliminate the majority of the non-$t\bar{t}$ background. Overall, a signal selection efficiency of 70.2\%, including the branching fractions of the considered fully-hadronic and semi-leptonic top pair decays, is achieved. As for the 500 GeV case, the dominant background channels are rejected at the 99.8\% level, resulting in an effective cross section for the remaining background of 73 fb.  

Simulated data points are generated by taking the ISR and luminosity-spectrum corrected top pair cross section at the desired energy to calculate the nominal number of events expected. The simulated number of signal events is determined on a random basis following a Gaussian distribution with the mean set to the nominal number of events and the standard deviation given by the square root of that number. With the same method, background events are added, using a constant cross section of 73~fb as discussed above. It is assumed that the nominal background contribution is well known both from theory and from measurements below threshold, so that the nominal number of background events can be subtracted from the signal, leaving just the statistical variations on top of the signal data with its own statistical uncertainty.

\begin{figure}
\centering
\includegraphics[width=0.99\columnwidth]{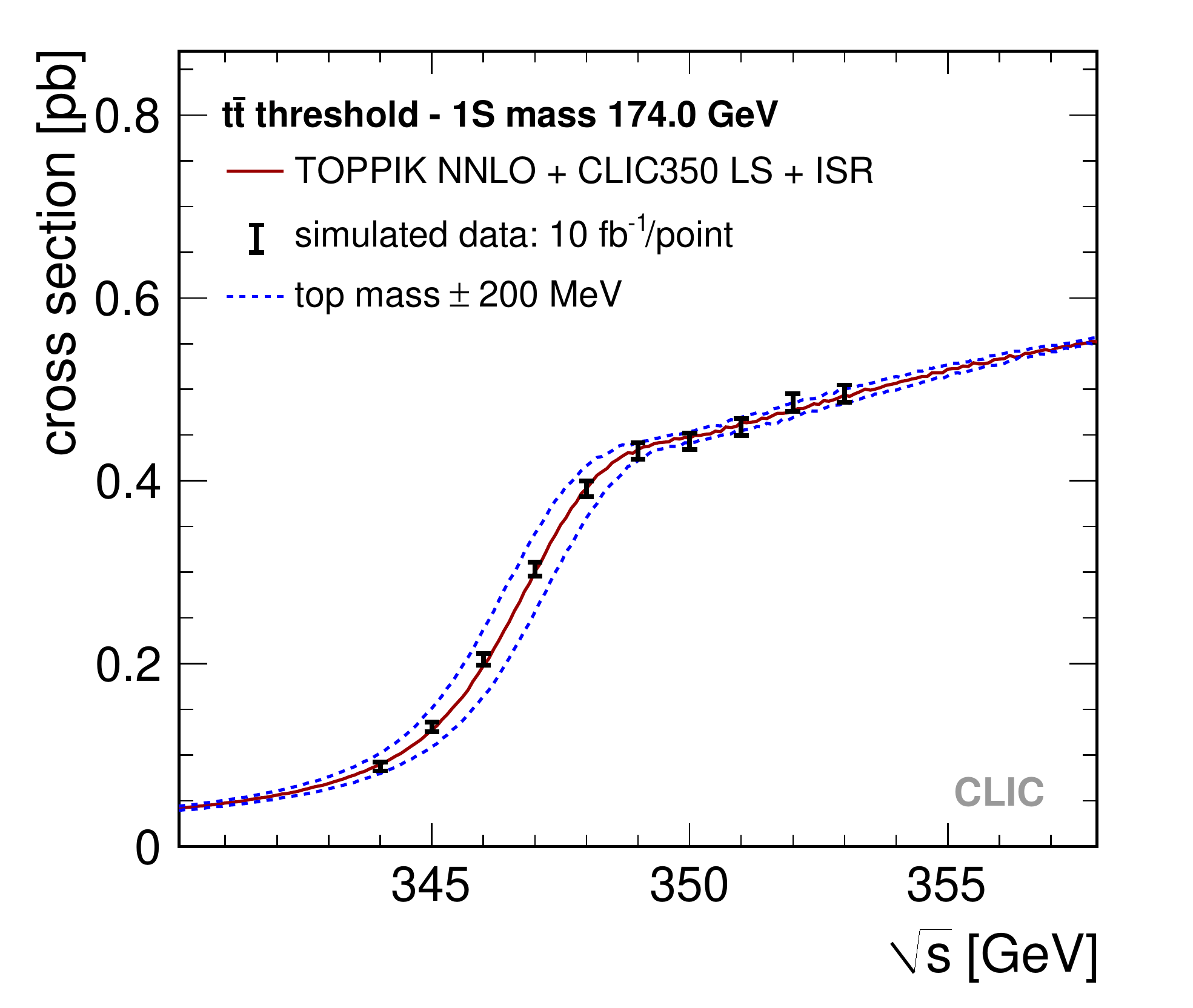}
\caption{Background-subtracted simulated cross section measurements for 10~fb$^{-1}$ per data point, together with the cross section for the generator mass of 174~GeV as well as for a shift in mass of $\pm$200~MeV.  \label{fig:ThresholdData}}
\end{figure}

Figure \ref{fig:ThresholdData} shows the ten simulated data points for CLIC with an integrated luminosity of 10~fb$^{-1}$ at each point. To illustrate the sensitivity of the data to the top quark mass, the threshold behavior for a shift in mass of $\pm$200 MeV is also shown in the figure.

\subsection{Measurement of the top mass and $\alpha_s$}

\begin{figure}
\centering
\includegraphics[width=0.99\columnwidth]{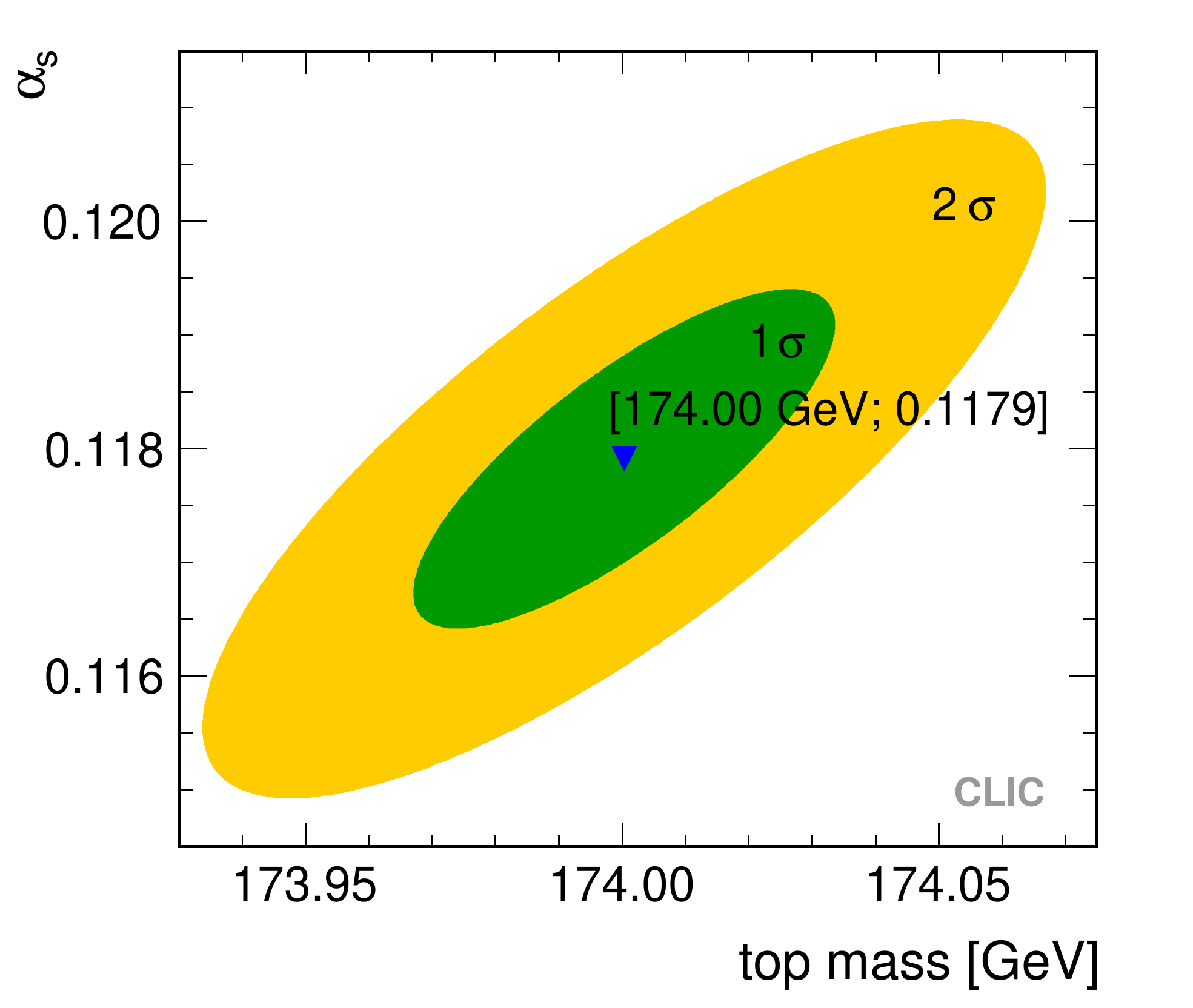}
\caption{Expected statistical errors from a simultaneous fit of the top mass and the strong coupling constant, showing the correlation of the two variables and the achieved precision. \label{fig:ThresholdResults}}
\end{figure}

The 1S mass of the top quark and the strong coupling constant are extracted simultaneously with a two-dimensional template fit. During the fitting procedure, the simulated data points are compared with calculated cross sections (``templates'') for a grid of different mass and strong coupling values, generated with step sizes of 50 MeV and 0.0007 for $m_t$ and $\alpha_s$, respectively. The fit results are then given by the minimum of a two-dimensional parabolic fit to the $\chi^2$ distribution of the different templates in the  $m_t$, $\alpha_s$ plane. The expected statistical uncertainty of these parameters from a threshold scan is taken from the standard deviation of the measured mass in 5000 trials with different simulated data points. The results are illustrated in Figure \ref{fig:ThresholdResults}, which shows the clear correlation between the two parameters, and also demonstrates that the fit itself does not introduce a bias on the results.

At this stage of the analysis, the systematic error due to theory uncertainties is included,  taken as an overall normalization uncertainty of the calculated cross section. Here, two levels are considered: A normalization uncertainty of 3\%, assumed as a reasonably conservative estimate of current theory uncertainties \cite{Hoang:2011it}, and an uncertainty of 1\% optimistically assumed to be achievable with additional theoretical work in time for experiments at linear colliders. 

The full results, including the theory uncertainty, are giv\-en in Table \ref{tab:ThresholdResults}.

\begin{table}
\centering
\begin{tabular}{l|c}
\hline
\multicolumn{2}{c}{1S top mass and $\alpha_s$ combined 2D fit}\\
\hline
 $m_t$ stat. error &  34~MeV\\
 $m_t$ theory syst. (1\%/3\%) &  5~MeV / 8~MeV\\
 $\alpha_s$ stat. error & 0.0009\\
 $\alpha_s$ theory syst. (1\%/3\%) & 0.0008 / 0.0022 \\
\hline
\end{tabular}

\caption{Summary of the 2D simultaneous top mass and $\alpha_s$ determination with a threshold scan at CLIC for 10 points with a total integrated luminosity of 100 fb$^{-1}$. \label{tab:ThresholdResults}}

\end{table}

\subsubsection{Alternative scenarios}

In addition to the two dimensional fit with 10 data points, other running and analysis scenarios are considered. When taking the strong coupling constant as an external input to the fit, the statistical uncertainty on the top quark mass reduces to 21 MeV. However, an additional systematic uncertainty of 20 MeV on the mass arises from the precision of $\alpha_s$, when assuming the current uncertainty of the world average of 0.0007 \cite{Bethke:2012jm}. In addition, the theory uncertainty of the overall normalization of the cross section is in that case fully absorbed by an uncertainty of the top mass, which amounts to 18 MeV and 56 MeV when assuming a 1\% and a 3\% uncertainty, respectively. Since the precision for the strong coupling constant achieved in the combined fit is comparable to the current uncertainty of the world average, there is no substantial benefit in using the strong coupling as external input unless the precision improves considerably by the time experiments are performed at linear colliders.

Since the sensitivity to the top mass is mainly provided by the region in which the cross section rises steeply a measurement is also possible with a reduced data set consisting of just the first six points in the range from 344 GeV to 349 GeV, with a combined integrated luminosity of 60 fb$^{-1}$. In this scenario, the statistical precision of the top mass is reduced by approximately 20\% to 40 MeV, with slightly reduced theory uncertainties. The uncertainty of the strong coupling on the other hand increases by 45\% to 0.0013, with unchanged theory systematics. This stronger increase in the error of the strong coupling constant compared to the top quark mass originates from the fact that the cross section above threshold, which is not measured in the six point scan, is particularly sensitive to the coupling constant.

\subsubsection{Additional systematic errors}

In addition to the theory normalization uncertainty other potential sources for systematic errors are studied. 

A potential dependence of the result on the choice of energy values for the scan in relation to the top mass was excluded by shifting the measurement points to higher energies by 0.5 GeV  without a change in the determined mass and $\alpha_s$ values. Thus, the precision expected from the LHC will be sufficient to determine the range of the threshold scan.

The precise knowledge of the non-top background after event selection is crucial for the measurement of the signal cross section. The effect of an imperfect non-top physics background description is studied by subtracting 5\% and 10\% too little or too much background before the fit. The 5\% variation results in a 18~MeV shift in the top mass and 0.0007 in $\alpha_s$, corresponding to approximately two thirds of the statistical uncertainty on the top mass and close to the statistical uncertainty on $\alpha_s$. The 10\% variation leads to a shift of twice the size for both values, but also significantly reduces the stability of the template fit. This shows that an understanding of the background contamination at the level of 5\% or better is important to keep systematic effects well below the statistical uncertainties. One additional data point taken below the threshold will provide this required precision. 

In addition to these analysis-related uncertainties, also machine-related uncertainties, such as the knowledge of the center-of-mass energy of the collider and the shape of the luminosity spectrum are relevant for this study. Previous experience at LEP \cite{Assmann:2004gc, Abbiendi:2004zw} and studies in the context of the ILC \cite{Boogert:2009ir, Hinze:2005kh} suggests that a precision of 10$^{-4}$ on the center-of-mass energy is readily achievable given the high available integrated luminosity at each data point, resulting in systematics  below the statistical errors of the top mass. The knowledge of the luminosity spectrum is very important for the correct description of the signal cross section, and thus also for the precision of the template fit. A full investigation has not yet been performed, but a simplified first study indicates that a 20\% uncertainty of the RMS width of the main luminosity peak results in top mass uncertainties of approximately 75~MeV, far in excess of the statistical uncertainties. Realistic studies of the expected uncertainties of the shape of the luminosity spectrum at CLIC are currently under way, but are beyond the scope of the present study. Overall, a good understanding of the luminosity spectrum of the accelerator is crucial to fully profit from the  precision achievable in a threshold scan. 

Taking these systematic studies together, it is to be expected that a threshold scan with 100 fb$^{-1}$ to determine the top mass at CLIC will be systematics limited, resulting in a total uncertainty below 100 MeV. 

\subsection{The impact of the luminosity spectrum: A threshold scan at ILC}

\begin{figure}
\centering
\includegraphics[width=0.99\columnwidth]{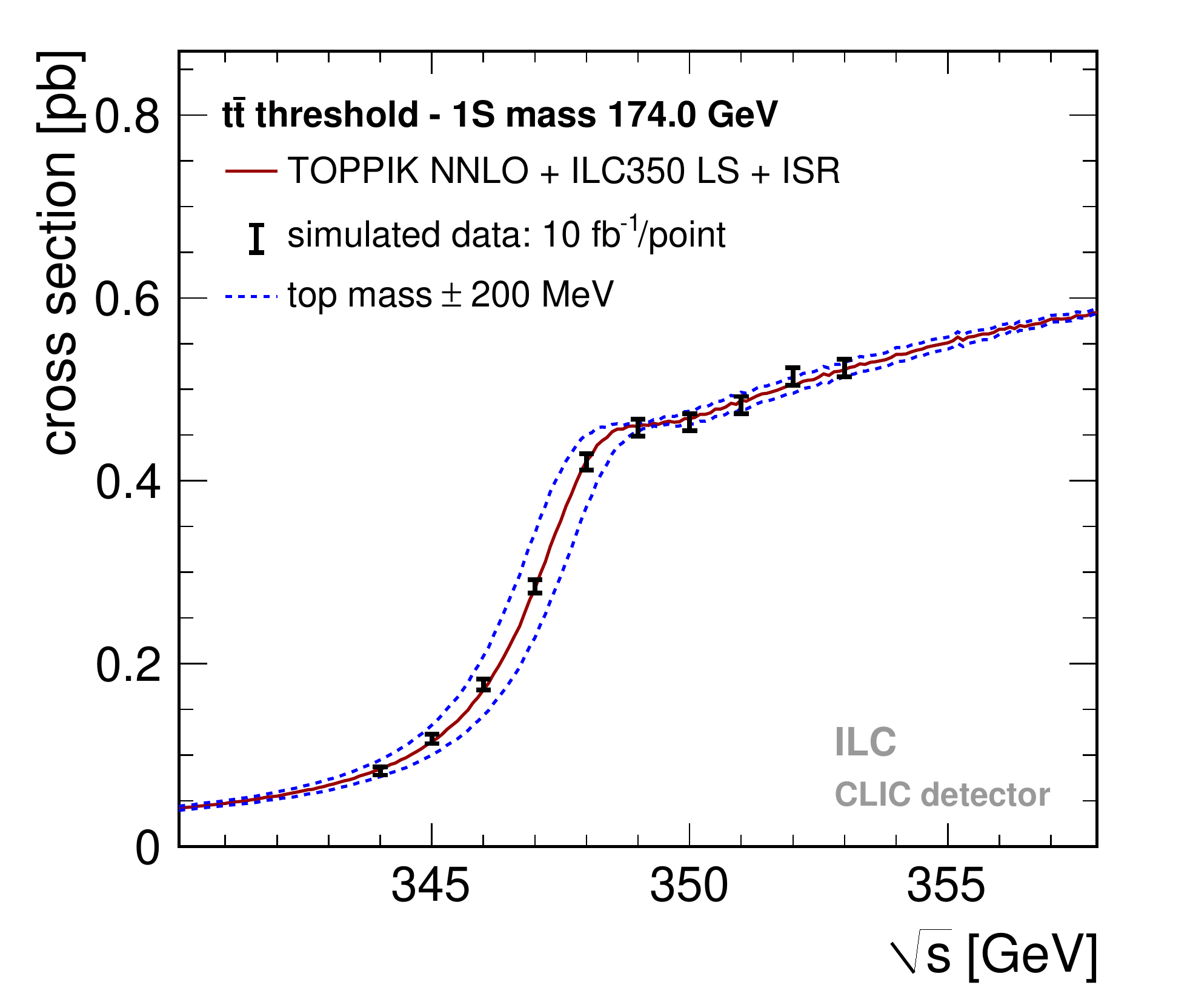}
\caption{Background-subtracted simulated cross section measurements with the ILC luminosity spectrum for 10~fb$^{-1}$ per data point, together with the cross section for the generator mass of 174~GeV as well as for a shift in mass of $\pm$200~MeV.  \label{fig:ThresholdDataILC}}
\end{figure}

For the experiments, one of the most relevant differences between CLIC and the ILC are the differences in the luminosity spectrum of the two machines. The influence of this difference is studied here by repeating the analysis using the ILC luminosity spectrum. The ILC spectrum is characterized by a narrower main luminosity peak, and a slight increase of the fraction of the total luminosity available in the top 1\% of the energy. As for the CLIC analysis, an integrated luminosity of 10 fb$^{-1}$ per point is assumed. Figure \ref{fig:ThresholdDataILC} shows simulated data points for a threshold scan at ILC. Compared to the CLIC case shown in Figure \ref{fig:ThresholdData}, the cross section rises faster due to the sharper main luminosity peak at the ILC. This faster rise of the cross section is expected to lead to somewhat reduced statistical uncertainties on the top mass for a given integrated luminosity due to increased differences between different mass hypotheses in the threshold region. 

For the generation of data points with the ILC luminosity spectrum, the signal selection efficiencies and the residual background contribution are determined with the CLIC\_ILD detector concept. While there are some differences between this detector concept and the ones developed for the ILC, it is not expected that this will have a sizeable impact on the efficiencies in the present study.

\begin{figure}
\centering
\includegraphics[width=0.99\columnwidth]{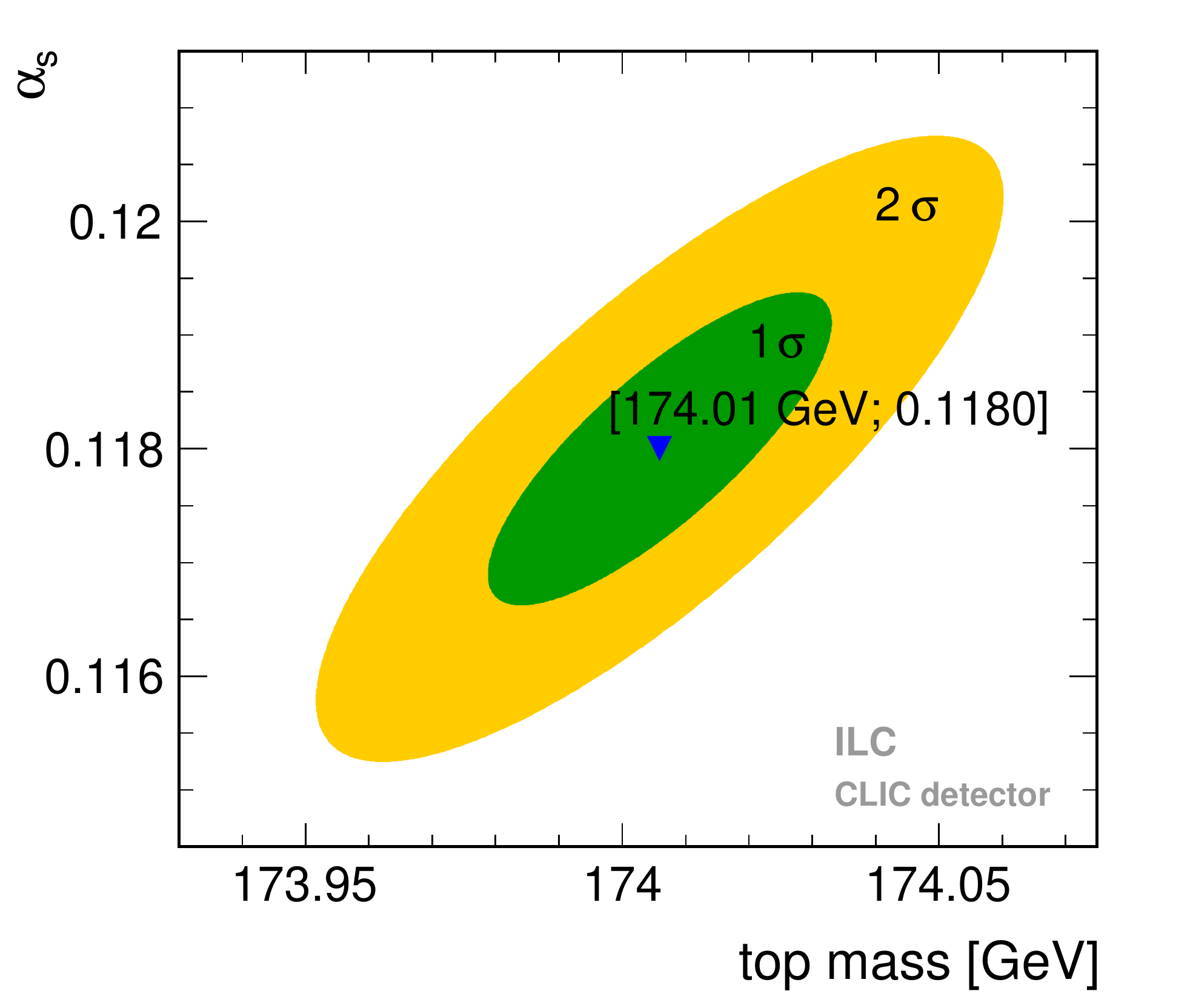}
\caption{Expected statistical errors from a simultaneous fit of the top mass and the strong coupling constant using the ILC luminosity spectrum, showing the correlation of the two variables and the achieved precision. \label{fig:ThresholdResultsILC}}
\end{figure}

\begin{table}
\centering
\begin{tabular}{l|c}
\hline
\multicolumn{2}{c}{1S top mass and $\alpha_s$ combined 2D fit}\\
\hline
 $m_t$ stat. error &  27~MeV\\
 $m_t$ theory syst. (1\%/3\%) &  5~MeV / 9~MeV\\
 $\alpha_s$ stat. error & 0.0008\\
 $\alpha_s$ theory syst. (1\%/3\%) & 0.0007 / 0.0022 \\
\hline
\end{tabular}

\caption{Summary of the 2D simultaneous top mass and $\alpha_s$ determination with a threshold scan at ILC for 10 points with a total integrated luminosity of 100 fb$^{-1}$. Event selection and background rejection from CLIC\_ILD is used.\label{tab:ThresholdResultsILC}}
\end{table}

Figure \ref{fig:ThresholdDataILC} and Table \ref{tab:ThresholdResultsILC} summarize the results of the combined extraction of the 1S top mass and the strong coupling constant at ILC. As expected, the statistical uncertainties are reduced compared to a threshold scan at CLIC, with a 20\% reduction of the uncertainty of the mass and a 10\% reduction of the uncertainty of $\alpha_s$. The theory systematics as well as other systematic uncertainties studied here are unchanged compared to those at CLIC. Thus, the difference in statistical precision provided by the two different collider concepts does not result in a significant difference of the overall precision of the top mass measurement in a threshold scan.

\section{Conclusions}

A linear $e^+e^-$ collider based on CLIC technology provides the capabilities for a precise measurement of the mass of the top quark both at and above threshold. We have studied the expected precision obtainable in top pair production events with a scan around the threshold and with the direct reconstruction of the invariant mass of the top decay products at an energy of 500 GeV, each assuming a total integrated luminosity of 100 fb$^{-1}$. The studies have been performed with realistic GEANT4-based detector simulations including physics and machine-related backgrounds using full particle flow event reconstruction. 

Above threshold, the mass of the top quark, here defined as the invariant mass of the decay products, can be measured with a statistical precision of 80 MeV combining fully-hadronic and semi-leptonic top pair decays. Systematic uncertainties originating from the jet energy scale can be controlled to a similar level using the direct reconstruction of the $W$ bosons in the top pair decays and $Z$ decays to $b\bar{b}$ from other sources. Since the measurement of the invariant mass is interpreted in the context of the top mass definition provided by the event generator PYTHIA, there are additional, potentially sizeable theoretical uncertainties when translating the result into theoretically well-defined mass schemes, which are not included in the quoted uncertainty. 

In a threshold scan, the top mass can be determined in a theoretically well defined way, here using the 1S mass, with a statistical precision of 34 MeV together with the strong coupling constant, which is determined with a statistical uncertainty of 0.0009. The theory uncertainty, incorporated as an overall normalization uncertainty of the cross section, is substantially smaller than the statistical error on the mass, and comparable to or larger than the statistical error on the strong coupling. Additional systematic uncertainties from the beam energy, from the luminosity spectrum and from the background subtraction are comparable or smaller than the statistical uncertainty on the mass, resulting in a total uncertainty of the top mass below 100 MeV in a theoretically well-defined mass scheme. The differences in luminosity spectrum between CLIC and ILC only lead to small differences in the statistical precision, far below the expected systematic uncertainties, demonstrating that also a collider based on CLIC technology is well suited for precision measurements in threshold scans.

In conclusion, these studies confirm the expectation that a linear $e^+e^-$ collider will be capable of measuring the mass of the top quark at the 100 MeV level, substantially beyond the precision expected at the LHC. However, for measurements above threshold the interpretation of the measured invariant mass currently still incurs sizeable theoretical uncertainties, requiring substantial advances in the theoretical understanding to utilize the experimental precision. The results demonstrate the benefits of the well-controlled luminosity spectra that allow the use of energy constraints and shows the possibilities provided by the high-resolution detector systems being developed for linear colliders.

\begin{acknowledgements}
We would like to thank Thomas Teubner and Andre Hoang for providing the TOPPIK code and for help with its use. The work presented in this paper has been carried out in the framework of the CLIC detector and physics study. It was supported in part by the DFG cluster of excellence ``Origin and Structure of the Universe'' of Germany. 
\end{acknowledgements}

\bibliographystyle{spphys}
\bibliography{ttbar}

\clearpage
\end{document}